\documentclass[aps,pra,twocolumn]{revtex4}
\usepackage{graphicx}
\usepackage{bm,dsfont}

\usepackage{amsmath, amsthm, amssymb,color}

\usepackage{euscript}

   \newtheorem{Theorem}{Theorem}

\newtheorem{definition}{Definition}
\newtheorem{proposition}{Proposition}
\newtheorem{corollary}{Corollary}
\newtheorem{lemma}{Lemma}

\def\bra#1{\mathinner{\langle{#1}|}}
\def\ket#1{\mathinner{|{#1}\rangle}}
\def\braket#1{\mathinner{\langle{#1}\rangle}}

{\catcode`\|=\active
 \gdef\Braket#1{\begingroup
\mathcode`\|32768\let|\BraVert\left<{#1}\right>\endgroup}}
\def\BraVert{\egroup\,\mid\,\bgroup}

\DeclareMathOperator{\tr}{tr}

\def\be{\begin{equation}}
\def\ee{\end{equation}}
\def\bes{\begin{eqnarray}}
\def\ees{\end{eqnarray}}
\def\bay{\begin{array}}
\def\ear{\end{array}}

\def\6{\langle}
\def\9{\rangle}
\def\tr{{\rm tr}\,}

\def\3{\mbox{$3\over2$}}
\def\1{\mathds{1}}

\def\eL{\EuScript{L}}

\def\openone{\1}
\newcommand{\mG}{{\mathcal{G}}}
\newcommand{\mGr}{{{\mathcal{G}}_{res}}}

\begin{document}
\preprint{}
\title{Discord and quantum computational resources}
\author{Aharon Brodutch}
\affiliation {Institute for Quantum Computing and department of Physics \& Astronomy, University of Waterloo, Canada }
\email{aharon.brodutch@uwaterloo.ca}
\date{\today}
\begin{abstract}
Discordant states  appear in a large number of quantum phenomena and seem to be a good   indicator  of  divergence from classicality.  While there is evidence that they are   essential for a quantum algorithm to have an advantage over a classical one,  their precise role is unclear.  We examine the  role of discord in quantum algorithms  using the paradigmatic framework of \emph{restricted distributed quantum gates} and  show  that manipulating  discordant states  using local operations has an associated cost in terms of entanglement and communication resources.  Changing discord reduces the total  correlations and reversible  operations on discordant states usually require non-local resources.  Discord alone is, however,  not enough to determine the need for entanglement. A more  general  type  of similar quantities, which we call $K$-discord, is introduced as a further constraint on  the kinds of operations that can be performed without entanglement resources.  

\end{abstract}
\maketitle

\section{Introduction}

In recent years quantum discord and similar quantities have become an actively studied topic. There is   mounting evidence that discordant states play an essential role in a wide variety of quantum phenomena (see \cite{RMP} and references therein).  In many cases  discord seems to  be  a more natural measure of quantum correlations than entanglement \cite{arXiv:0709.0548,RMP}  and has sometimes been hailed as an essential resource for mixed state quantum computation \cite{arXiv:1109.5549}.  The term \emph{resource} is, however, debatable, mainly due to the fact that  very few limitations have been found on the creation and manipulation of discord. Most notably, unlike entanglement, discord can be created and increased using local operations \cite{RMP}. 

Nevertheless,  there is strong evidence  that discordant states  play a role in mixed state  protocols and  discord is in many ways an indication of the divergence from classicality \cite{RMP,arXiv:1109.5549,discordresource,MileNature}. In the context of quantum algorithms there is  interest in identifying ``the elusive source of quantum speedup'' \cite{elusive}. Recent results indicate that discordant states play an essential role in quantum algorithms that display an advantage over classical ones  \cite{RMP,arXiv:0709.0548,eastin}.

Discord shares many properties with  pure state entanglement measures \cite{RMP}.  These similarities  and the inability to extend some results regarding entanglement in pure state quantum computation  to mixed states \cite{JozsaLinden, Vidal,VdN} make discord one possible candidate  for the source of computational speedup \cite{arXiv:quant-ph/0110029}. There are, however obvious differences between discord in mixed states and entanglement in pure states. The most significant is that discord is not known to be monotonic under a relevant  class of operations.  

 Another approach for  identifying the source of the quantum advantage is to attribute it  to quantum gates \cite{PhysRevLett.83.1054,Meznaric}. One way to reconcile the   two pictures (gates vs. states) is to look at the relation between the sets of  input and output states. It is also possible to bring entanglement back into the picture by examining a \emph{local operations and classical communication} (LOCC) implementation of the relevant quantum gates. Indeed under reasonable assumptions an LOCC protocol that does not require ancillary  entanglement can be simulated efficiently on a classical computer\cite{NielsenHarrow}.  

In \cite{BTgates}   we introduced \emph{restricted, distributed gates} to study the question: \emph{when is  a reversible operation effectively non-classical?} in terms of entanglement resources.    The result was a  relation between discord in the set of input/output states and entanglement. It was however, limited to  cases where only rank-1 measurements are applicable.  The original definition of discord  naturally selects rank-1  measurements since they maximize information gain  \cite{RMP, OllivierZurek01,HendersonVedral01,datta}. However, when  discussing  LOCC protocols it is often useful to consider  measurements that reveal less information and are consequently less disturbing. In this regard one should define a more general version of discord. Similar issues were recently discussed  in  \cite{arXiv:1202.1598,LuoFu2012} where orthogonal  projective measurements of different ranks were considered.

Here  we extend the results of \cite{BTgates} and  tackle a number of issues described above. First we show that changing discord using local operations has an associated cost in terms  of mutual information.  Next we define a  discord-like quantity that  takes into account  more general measurements.   We use this quantity to extend the results of \cite{BTgates} to  more general types of states including those used in NMR  mixed state quantum computation.  The results indicate that in the context of reversible operations discord acts more like an obstacle then a resource. Non-local  resources are required to \emph{change} discord rather then just to increase it.  These results, as well as the tools used to obtain them, are useful for a conceptual understanding of the quantum advantage and can be extended to answer more general questions. 
\vskip 7pt
%%%%%%%%%%%%%%%%%%%%%%%%%%%%%%%%%%%%%%%%%%%%%%%
\subsection{Notation}
The state of a quantum system shared by two parties, Alice and Bob, is denoted by $\rho_{AB}$ with reduced states  $\rho_A=\tr_B[\rho_{AB}]$ and $\rho_B=\tr_A[\rho_{AB}]$.
Operations on Alice's side are described by their Kraus operators $\{M_a\}$. In principle any operation can be called  a measurement with  POVM elements $\{E_a=M_a^\dagger M_a\}$ and a ``classical'' measurement outcome  corresponding  to each  term.  We use the term \emph{measurement} whenever an operation has at least one POVM element   which is not proportional to the identity $\openone$.   A measurement is  rank-1  if and only if all the POVM elements $\{E_a\}$ are rank-1. 

The   probabilities for the classical outcomes are given by  $p_a=\tr(E_a\rho_{A})$. The resulting (conditional) state is  $\rho_{AB|a}=M_a\rho_{AB}M_a^\dagger$. The conditional state on Bob's side is $\rho_{B|a}=\tr_A(\rho_{AB|a})=\tr_A(E_a\rho_{AB})$.  If we discard the classical outcomes, the resulting (average) state is given by  $\rho_{AB}'=\sum_aM_a\rho_{AB}M^\dagger_a$.  This  final state can include any  ancillary systems used on Alice's side, thus $d_A=dim(\rho_A)$ is not necessarily the same as $d_A'=dim(\rho_A')$. In principle it is possible to encode the results in orthogonal states, for example  $\rho_{AB}'=\sum_a\ket{a}\bra{a}\otimes\rho_{B|a}$ or $\rho_{AB}'=\sum_a\ket{a}\bra{a}_{A_L}\otimes\rho_{ A_RB|a}$ where the Hilbert space  $\mathcal{H}_A=\mathcal{H}_{A_L}\otimes\mathcal{H}_{A_R}$ and $\braket{a|b}=\delta_{ab}$.  For simplicity of notation we sometimes use $\Lambda_A$ to denote an operation  on Alice's  side such that $\Lambda_A(\cdot)=\sum_aM_a\cdot M_a$. $\Lambda$ without a subscript represents a local operation on both sides.

%%%%%%%%%%%%%%%%%%%%%%%%%%%%%%%%%%%%5
\section{Discord and local operations}
\subsection{Quantum correlations} 
 The total correlation in a bipartite quantum system is  given by the mutual information $I(A:B)=S(\rho_{AB}||\rho_{A}\otimes\rho_B)$ where $S(\rho||\tau)=\tr[\rho\log\rho-\rho\log\tau]$ is the quantum relative entropy.  

Discord was originally defined via the  probabilities for the outcomes of a measurement \cite{OllivierZurek01,HendersonVedral01}: 
\begin{equation}\label{discord}
D(B|A)=I(A:B)-\max_{\{E_a\}}[S(B)-\sum_ap_{a}S(\rho_{B|a})]
\end{equation}
 The maximization  depends only on the POVM elements $\{E_a\}$  and  naturally selects rank-1 measurements \cite{datta,RMP}. Interpreting the last term as the classical part of the correlations, $J(B|A)=\max_{\{E_a\}}[S(B)-\sum_ap_{a}S(\rho_{B|a})]$,  gives
\begin{equation}\label{MICD}
I(B|A)=J(A:B)+D(B|A).
\end{equation}
Discord is then the quantum part of the correlations \cite{OllivierZurek01,HendersonVedral01,BrodutchModi}. 

We can also define quantum discord as the minimal change in mutual information after a  rank-1  measurement \cite{BTdiscord}:
\begin{equation}\label{discordmi}
 D(B|A)=\min_{\{\bar M_a\}}[I(AB)-I(A'B')].
\end{equation}
 The bar indicates that  the corresponding POVM elements $\{\bar E_a  =\bar M^\dagger_a\bar M_a\}$  are rank-1. The two definitions, Eqs. (\ref{discord}) and (\ref{discordmi}), are equivalent.

 Zero discord states are called \emph{classical} \cite{RMP}. A state is classical if an only if it has the form  

\begin{equation}\label{eqclassical}
\rho_{AB}=\sum_{a}p_{a}\Pi_a\otimes\rho_{B|a},
\end{equation}
where  $\{\Pi_a\}$ is a set of  orthogonal projectors  \cite{OllivierZurek01}. These  states have only classical correlations. For classical states it is possible to find  a local  rank-1 measurement that does not induce any loss of information. The POVM elements of this measurement are $\{\Pi_a\}$.

 Discord, classical correlations and mutual information  are invariant under local unitary operations.   Mutual information and classical correlations are also non-increasing under local operations \cite{HendersonVedral01}. Discord on the other hand can both increase and decrease under local operations \cite{RMP}.  Both discord and classical correlations are  not symmetric under the interchange of the subsystems. Here  we always consider  the discord under a measurement on Alice's subsystem. 
\vskip 7pt

\subsection{Changing discord with  local operations}
To show that changing discord implies the loss of correlations we  will use  a special case of Petz's theorem \cite{petz,arXiv:quant-ph/0304007}  regarding the reversibility of completely positive trace preserving (CPTP) operations.

{\lemma \label{petz} Given two states $\rho_{AB}$ and $\tau_{AB}=\tau_A\otimes\tau_B$, where $S(\rho_{AB}||\tau_{AB})<\infty$, and a local (CPTP) operation $\Lambda$, the  equality $S[\rho_{AB}||\tau_{AB}]=S[\Lambda_A(\rho_{AB})||\Lambda(\tau_{AB})]$ holds if and only if there exists a local operation $\Lambda^*$ such that $\Lambda^*[\Lambda_A(\tau_{AB})]=\tau_{AB}$ and $\Lambda^*[\Lambda(\rho_{AB})]=\rho_{AB}$  \cite{arXiv:0707.0848}. 
}

Taking   $\tau_{AB}=\rho_A\otimes\rho_B$  we get a corollary: $I[\rho_{AB}]=I[\Lambda(\rho_{AB})]$  if and only if $\Lambda$ can  be reversed locally.   We now present our first result.

 {\Theorem If a local operation $\Lambda(\rho_{AB})=\rho_{AB}'$ changes discord,  $D(B|A)\ne D(B'|A')$, it also decreases mutual information, $I(\rho_{AB})>I(\rho_{AB}')$. \label{theoremMI}}
 
 Alternatively, using the fact that local operations cannot increase mutual information, the theorem reads: 
 
 {\it For a local operation, $\Lambda$, we have $I[\rho_{AB}]=I[\Lambda(\rho_{AB})]\Rightarrow D(B|A)=D(B'|A')$.} 

{\proof  First we show that discord cannot be decreased without affecting mutual information. Neither mutual information, $I(A:B)$, nor classical correlation, $J(B|A)$, can increase under local operations \cite{HendersonVedral01}. Using eq. \eqref{MICD} we see that decreasing discord reduces mutual information by at least the same amount.

To show that discord cannot be  increased we use lemma \ref{petz}.   The operation   $\Lambda$ must be  reversible if mutual information does not change. If $\Lambda$  increases discord without changing mutual information then its reverse can decrease discord without changing mutual information, violating the first part of the  proof.  \qed}  

Note that  this proof is valid for local operations on both Alice and Bob's side.

	Theorem \ref{theoremMI} implies that a local operation that {\it changes}  discord   results in some  loss of information. In the case where the initial state is unknown this loss of information will usually affect reversibility even when we allow classical communication.  In what follows we will show that reversible  operations taking  sets of initial states to final states require  non-local resources if discord changes. This is a consequence of  the effects of measurements on discord and mutual information.  

\section{Discord and entanglement resources}
\subsection{Restricted distributed gates} 

To illustrate some of the implications of theorem \ref{theoremMI} we extend  the \emph{restricted distributed gates} paradigm  first introduced in \cite{BTgates}:  Alice and Bob may use arbitrary LOCC operations to implement a reversible quantum gate $\mG$  on a restricted set of input states, $\eL$.  This  corresponds to  realistic situations where the input states are very seldom arbitrary.  In particular it corresponds to the case where at each stage of the computation the system is separable as in some mixed state algorithms \cite{RMP}.  

Restricted distributed gates may prove to be useful for some quantum computing tasks. However, here we use this paradigm as a theoretical tool for studying `standard' quantum information processing scenarios, in particular the quantum circuit model. With this in mind we restrict ourselves to reversible operations.

{\definition A quantum CPTP  operation, $\mG$, is called reversible on the set of states $\eL$ if and only if  there exists an inverse CPTP operation   $\mG^{-1}$ such that $\mG^{-1}[\mG(\rho^i_{AB})]=\rho^i_{AB}$ for all $\rho^i_{AB}\in\eL$.}

A unitary gate is reversible for any input set. More generally  irreversibility  is a consequence of information loss and reversibility  can be related to error correction.

{\definition Given a reversible operation, $\mG$, on a set of bipartite states, $\eL$, we define the  distributed, restricted gate, $\mGr$ on  $\eL=\{\rho^i_{AB}\}$  as  a CPTP, LOCC  operation  with  
\be
\mGr(\rho^i_{AB})=\mG(\rho^i_{AB})
\ee
 for all states $\rho_{AB}^i\in \eL$.
}

In general $\mGr(\rho^x_{AB})\ne\mG(\rho^x_{AB})$ when $\rho^x_{AB}\notin\eL$.  Since the operations are linear we can assume  $\eL$ is  convex without loosing generality. 

When there are no  entangled ancillary systems we  can, without loosing generality, describe the implementation of $\mGr$ as a set of   local CPTP operations $\Lambda_\mu$ at each stage followed by classical communication from Alice to Bob $C_{A\to B }$ or Bob to Alice   $C_{B\to A }$. The classical information is then encoded as part of the local quantum state before the next  step. It  is retained throughout the operation and discarded  (traced out) at the end,
\be\label{resgate}
 \mGr(\rho_{AB})=\tr_{cl}\;\Lambda_{A_n}\;C_{B\to A}\;...\;C_{A \to B}\;\Lambda_{A_1}(\rho_{AB}).
\ee
Without loss of generality we always assume Alice goes first. 

 Reversibility of $\mGr$  requires  $S(\mG[\rho_{AB}^i]||\mG[\rho_{AB}^j])=S(\rho_{AB}^i||\rho_{AB}^j)$ for  all $\rho^i_{AB},\rho^j_{AB}\in\eL$. Using the above structure this requires that at any stage the relative entropy remains constant, in particular  $S(\Lambda_{A_1}[\rho_{AB}^i]||\Lambda_{A_1}[\rho_{AB}^j])=S(\rho_{AB}^i||\rho_{AB}^j)$.   Note that $G^{-1}$ is not restricted to LOCC.

 If any communication is necessary during the protocol then the  first step, $\Lambda_{A_1}$,  should involve a measurement whereby Alice can gather  some information about Bob's (conditional) state.  A measurement that reveals  maximal information (and maximizes the last term in Eq. \eqref{discord}) would in general change discord and in most cases decrease the relative entropy between some states in the set of initial states.  To overcome this problem Alice can choose her first measurement to reveal less information.  In principle Alice and Bob may have some physical restriction on their  local measurements. We denote the set of allowed measurements  $\mathcal{S}^K$. In the most general case Alice and Bob can make any  measurement.  We denote the set of all measurements  $\mathcal{S}^2$ for a reason that will be apparent below.

We  now ask \emph{Given a set of measurements $\mathcal{S}^K$ can Alice and Bob implement the gate $\mGr$ without ancillary entanglement?} If the answer is yes we say that $\mGr$ is pseudo-classical for $\mathcal{S}^K$, otherwise we say it is non-local for  $\mathcal{S}^K$. $\mGr$ is fully non-local when it is non-local for $\mathcal{S}^2$. It is  fully local if it is pseudo-classical for the empty set. In the fully local case Alice and Bob do not need to communicate in order to implement   the gate.

\subsubsection*{Examples for $\mathcal{S}^K$}
The set of measurements $\mathcal{S}^K$ will generally depend on the physical scenario. One set that has been extensively studied  in the past is the set of all rank-1 orthogonal projective measurements or fully dephasing  channels $\mathcal{S}^\Pi$. The corresponding set of $\Pi$-classical states (see below for a definition of $K$-classical states) is the same as the set of zero discord (or classical) states in eq. \eqref{eqclassical}.

More general  are the sets of  all rank $r$ orthogonal projections, $\mathcal{S}^{R=r}$. Classicality under these sets of measurements  was studied in \cite{LuoFu2012}. While these sets have a simple mathematical structure it is not clear what is the  physical scenario where such restrictions apply. In \cite{arXiv:1202.1598} these sets were studied  from the perspective of local unitary operations with degenerate eigenvalues.

Other sets of interest are the sets of measurements with at least  $N$  outcomes (or linearly independent POVM elements) $\mathcal{S}^N$. The physical motivation here is a bit more clear since the number of outcomes is a property of the measurement device. A special case which has a clear physical significance is $N=2$. This is the set of all possible measurements since a measurement must have at least two outcomes. Some of our main results below  will be derived using this set of all measurements,  $\mathcal{S}^2$.  

%%%%%%%%%%%%%%%%%%%%%%%%%%%%%%%%%%%5

\subsection{K-classical states}

We now  define a notion of classical states  with respect to  a set of measurements $\mathcal{S}^K$. The relation between this definition and  the classicaity of distributed gates will become apparent  in the next sub-section.

{\definition A state is called $K$-classical if  $I[\rho_{AB}]=I[\Lambda_A(\rho_{AB})]$  for some measurement $\Lambda_A\in \mathcal{S}^K$.  A state is $K$-discordant if it is not $K$-classical. \label{defkclassical}}

 $K$-classical states can have non zero discord.   An example is a $3\times3$ discordant state which is a mixture of a  maximally entangled state, $1/\sqrt{2}[\ket{00}+\ket{11}]$ and a product state,  $\ket{22}$, such a state is  2-classical (classical with respect to the set of all measurements)  and entangled. 

Classicality under  $\mathcal{S}^N$ -the set of measurements with at least $N$ outcomes- is  related to reversibility under $\mathcal{S}^N$.

{\proposition A state $\rho_{AB}$ is $N$-classical if and only if there exists $\Lambda_A\in\mathcal{S}^N$ such that $\Lambda_A(\rho_{AB})=\rho_{AB}$}.

Proof follows directly from lemma \ref{petz}.

%{\definition A convex set of bipartite states, $\eL=\{\rho^i_{AB}\}$ is $K$-classical if and only if  all states in this set  are $K$-classical. }

%%%%%%%%%%%%%%%%%%%%%%%%%%%%%%%%%%%%%%%%%55

We use  the following necessary and sufficient condition for $K$-classicality to show  a relation between  $K$-discordant states, the change in discord, and  entanglement resources.  

{\lemma \label{lemmathermal} A state $\rho_{AB}$ is $K$-classical if  and only if there is a local measurement  $\Lambda_A\in S^K$  and a product state $\tau_{AB}=\tau_A\otimes\tau_B$, with $S(\rho_{AB}||\tau_{AB})<\infty$, such that  $S[\rho_{AB}||\tau_{AB}]=S[\Lambda_A(\rho_{AB})||\Lambda_A(\tau_{AB})]$.

\proof{  From lemma \ref{petz} the equality above holds if and only if there is  a local  reverse operation  operation $\Lambda_A^*$. This reversibility condition is also necessary and sufficient for $K$-classicality since a local operation which reverses $\Lambda_A$ on $\rho_{AB}$ will also reverse it on $\rho_A$  so $\Lambda_A$ preserves mutual information.  \qed}

\subsection{Operations that require entanglement}

Let us now examine the following general scenario: Alice and Bob are limited to local operations where all measurements are in the set $\mathcal{S}^K$. They implement a restricted gate $\mGr$ over a set of states $\eL$ that includes a product state $\tau_{AB}=\tau_A\otimes\tau_B$ and a $K$-discordant state $\rho_{AB}$ with   $S(\rho_{AB}||\tau_{AB})<\infty$.  What can we say about the non-local resources required for this gate?

  From Lemma \ref{lemmathermal} any measurement on Alice's side  $\Lambda_{A_1}\in S^K$,  will decrease the relative entropy and will  conflict with the reversibility condition.This leaves two options: (a) the implementation of $\mGr$ does not require any measurements in which case it is fully local or (b)  some ancillary entanglement is necessary for the  implementation,  so $\mGr$   is non-local for $\mathcal{S}^K$. 

In the special case where there is no restriction on the local measurements, $\mathcal{S}^K=\mathcal{S}^2$,  the restricted gate is either fully local (can be implemented without communication or entanglement) or fully non-local (requires non-local resources). In the fully local case we can appreciate theorem \ref{theoremMI}. Since the operation is fully local any change in discord would reduce mutual information. Consequently the operation would not be reversible locally and the relative entropy between $\rho^{AB}$ and $\tau^{AB}$ would decrease  (lemma \ref{lemmathermal}), violating    the (non-local)  reversibility condition. The above  proves  the following:

{\Theorem   A restricted, distributed  gate on an input set $\eL$ that includes a product state $\tau_{AB}=\tau_A\otimes\tau_B$  cannot be implemented  without entanglement resources if it changes  discord for any $2$-discordant state $\rho_{AB}\in \eL$, with $S(\rho_{AB}||\tau_{AB})<\infty$.}

The condition $S(\rho_{AB}||\tau_{AB})<\infty$ is always satisfied when $\tau_{AB}$ is the completely mixed state.  In this case the convex  set $\eL$ also includes  noisy states of the form  $\rho_{AB}=(1-N)\rho_{AB}^0+N\openone_{AB}/d_{AB}$. These are the standard states for  NMR quantum information processing  \cite{arXiv:quant-ph/0110029}. 

{\corollary   Take the noisy family of states  $\rho_{AB}=(1-N)\rho_{AB}^0+N\openone_{AB}/d_{AB}$ were $0\le N\le 1$ is a free parameter and $\rho_{AB}^0$  is $2$-discordant. A restricted gate on this family cannot be implemented  without entanglement resources if it changes the discord of $\rho^0_{AB}$. \label{C1} }

\vskip 7pt
\subsubsection*{Example}
  Consider  a distributed quantum  simulation of an NMR protocol that includes only a single factorized  input state $(1-N)\rho_{A}\otimes\rho_{B}+N\openone_{AB}/d_{AB}$  but varied amounts of noise $N$. Assume that at some point during the computation the state  becomes $2$-discordant across the relevant bi-partition. The next quantum gate that changes discord  will require some ancillary entanglement to simulate on  a distributed quantum computer. Simulate here means producing the exact output state.  The relation to classical simulation   is not immediate, but in the case where a fully local  implementation exists there is an efficient  classical algorithm for obtaining an arbitrarily good estimate of the possible measurement probabilities  at the end of the quantum  algorithm  \cite{NielsenHarrow}. 

\vskip 7pt

\section{Discussion and outlook} 

The results presented here are useful for tackling a number of questions regarding the role of discord and discordant states in various processes. Manipulating  discordant states has an associated resource  cost.  Any  change in  discord using local operations results in a loss of mutual information.

With respect to quantum computational resources, it is possible to regard entanglement as a resource by examining a distributed implementation of the relevant algorithm. In this picture there is a cost associated with manipulating discord. This cost is related to a more general sense of classicality then the one defined by zero discord. Classicality under the set of all local measurements, which we call $2$-classicality,   is the most sensitive indicator of the requirement for  entanglement resources. Changing discord in a noisy 2-discordant set requires non-local resources when the operation is reversible.  In light of theorem \ref{theoremMI} this is also a statement about decreasing mutual information. Either way discord should not be viewed as a \emph{resource} in this context. 

  $2$-discordent states provide an obstacle that can only be overcome using non-local resources.  More generally $K$-discordant states are an obstacle that can be overcome by either non-local resources or less disturbing measurements then those in $\mathcal{S}^K$. Each of these limitations has implications regarding   quantum resources.  One can also try to quantify the set of allowed measurements as a resource. 

This notion of discord as an obstacle was used to discuss the verification of non-local gates  in the cases of one way  communication  \cite{MileNature} and   qubits \cite{Almeida}. The  definition of $K$-discord allows a more general approach to this problem. Surprisingly variable depolarizing noise can help the verifier (Corollary \ref{C1}). 

\subsection{$K$-discord}

The  results are  qualitative: we showed  that changing discord has an associated resource cost  but we did not calculate this cost explicitly. A  quantitative relation between discord in a set of initial states  and the  entanglement and communication resources required to manipulate the states in this set is still missing.   Another missing parameter is  a quantitative measure of $K$-discord  defined in an operationally meaningful way. Apart from  vanishing  for $K$-classical states it should be directly related to the problem at hand.   One candidate is  the expression: $\inf_{\Lambda_A\in\mathcal{S}^K}\left\{I[\rho_{AB}]-I[\Lambda_A(\rho_{AB})]\right\}$.  In the case of $\mathcal{S}^{R=1}$ this is the standard quantum discord, eq. \eqref{discord}. For $R>1$ this expression was analyzed in \cite{LuoFu2012}. In many other cases and in particular for $\mathcal{S}^N$ this quantity may vanish for all states   \cite{privcom} and some meaningful normalization should be used. Alternatively the distance from the set of $K$-classical states $\mathcal{C}^K$ can be used eg. $\min_{\varrho_{AB}\in\mathcal{C^K}}S(\rho_{AB}||\varrho_{AB})$ in a similar way to other operationally meaningful quantities  \cite{RMP,arXiv:10292162,arXiv:0911.5417}.

$K$-discord can  also  be related to the thermal discord   (see \cite{RMP,BTdiscord,Zurek03})  using the distance to the completely mixed state eg. by taking the difference:  $S[\rho_{AB}||\openone_{AB}/d_{AB}]-S[\Lambda_A(\rho)||\Lambda_A(\openone_{AB}/d_{AB})]$, such a quantity may  play a role in finding the number of rounds (times the parties exchange information) required for extracting maximal work from a heat engine. It is  also  related to an open question regarding  the difference between the one-way and two-way quantum deficit \cite{RMP,arXiv:quant-ph/0112074}.

 A quantitative relation between discord and entanglement resources in restricted gates  may also answer practical questions regarding the use of restricted gates in place of standard reversible gates.  It would be interesting to find a useful algorithm for restricted gates that are simple to implement  compared to  their non-restricted counterparts. Moreover such a quantitative relation will   provide an operational approach to  a long sought-after resource theory for discord. Here discord indicates the need for other resources rather then playing the role of a resource per-se.

\emph{Acknowledgements} - I  thank  Eric Chitambar, Myungshik Kim, Kavan Modi, Marco Piani and Daniel Terno for useful discussions and comments.  This work was supported by  Industry Canada, NSERC  and CIFAR.

\end{document}